\documentclass[preprint,aps,floats,floatfix,nofootinbib,superscriptaddress]{revtex4}
\usepackage{graphicx} \usepackage{bm} \usepackage{epsfig}
\usepackage{pstricks}

\begin{document}

\title{Absorption Effects due to Spin in the Worldline Approach to Black Hole Dynamics}

\author{Rafael A. Porto}
\affiliation{Department of Physics, University of California, Santa Barbara, CA 93106}

\begin{abstract}
We generalize the effective point particle approach to black hole dynamics to include spin. In this approach dissipative effects are captured by degrees of freedom localized on the wordline. The absorptive properties of the black hole are determined by correlation functions which can be matched with the graviton absorption cross section in the long wavelength approximation. 
For rotating black holes, superradiance is responsible for the leading contribution. The effective theory is then used to predict the power loss due to spin in the dynamics of non--relativistic binary systems. An enhancement of three powers of the relative velocity is found with respect to the non--rotating case. Then we generalize the results to other type of constituents in the binary system, such as rotating neutron stars. Finally we compute the power loss absorbed by a test spinning black hole in a given spacetime background. 
\end{abstract}

\maketitle

\section{Introduction}

A new formalism, based on Effective Field Theory (EFT) techniques \cite{eft}, has been recently developed to systematically compute high order contributions to the gravitational potential and radiation for non--relativistic spinning binary systems within the Post--Newtonian (PN) framework  \cite{nrgr,nrgr4,nrgr3,nrgr4n}. Moreover, this new approach has led to new results at third Post-Newtonian order \cite{eih,nrgr5,comment,nrgrss,nrgrs2,srad}. 
Dissipative effects can be included as well within the EFT by introducing new degrees of freedom in the worldline action \cite{nrgr2}. Even though we lack a precise account for these internal degrees of freedom\footnote{Perhaps in the case of the BH, these new degrees of freedom could be associated with dynamics on the stretched horizon along the lines of the Membrane paradigm \cite{memb}}, it is possible to rely on the symmetries of the problem to construct an EFT to describe absorption. Therefore, the set of higher dimensional operators added to the worldline take the form \cite{nrgr2}
\begin{equation}
\label{eq:BHgrav}
S_{int}=-\int d\tau Q^E_{ab} E^{ab}-\int d\tau Q^B_{ab} B^{ab}+\cdots,
\end{equation} 
with $E^{ab}=e^a_\mu e^b_\nu E^{\mu\nu}$, $B^{ab}=e^a_\mu e^b_\nu
B^{\mu\nu}$, the electric and magnetic component of the Weyl tensor, and $e^a_\mu$, $a,b = 1 \ldots 3$, represents a local frame orthogonal to the four velocity $v^\mu$, which is parallel transported along the worldline. 
The new degrees of freedom are encapsulated in the quadrupole--like  terms $Q^{E,B}_{ab}$.  Within the EFT spirit it is possible to express physical observables in terms of correlation function of these new operators. In practice we will be able to match in the low frequency approximation for the Green's functions in known scenarios, such as the graviton absorption cross section for a rotating BH, and gain predictive power in those more complicated ones such as binary systems. In what follows we review the results for non--rotating binary BHs of Ref. \cite{nrgr2}, and afterwards extend the formalism to include effects due to spin. An enhancement of three powers of the relative velocity is found with respect to the non--rotating case. Then we generalize the formalism to include other type of constituents in the binary system, such as rotating neutron stars (NSs). For the case of the NSs we will be able to make predictions in terms of the graviton cross section given a theory of the internal structure of the NS. Finally we compute the power loss absorbed by a test spinning black hole in a given spacetime background. The last two results have not appeared in the literature until now.\\

We use $\hbar=c=1$ units. The Planck mass is given by $m_p=\frac{1}{\sqrt{32\pi G_N}}$, where $G_N$ stands for the Newton's constant. 

\section{Absorption effects for non--rotating binary Black Holes}

\begin{figure}[t!]
    \centering
    \includegraphics[width=6cm]{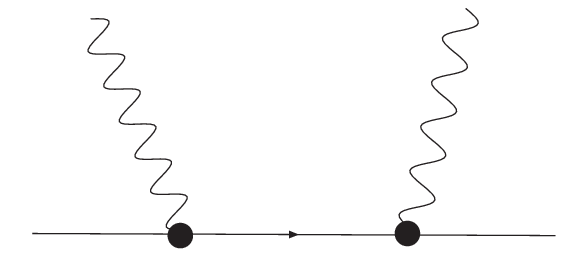}
\caption{Feynman diagram whose imaginary part gives the leading order contribution to the absorptive cross section. The dots correspond to insertions of leading multipole worldline operators.}
\label{subD} 
\end{figure}

Here we review the main results in \cite{nrgr2} which we will generalize to include spin in the next section. 
Using (\ref{eq:BHgrav}), the graviton absorption cross section is given, via the optical theorem, by the imaginary part of diagram Fig. \ref{subD} \cite{nrgr2}
\begin{eqnarray}
\sigma^{eft}_{abs}(\omega)&=& 2~\mbox{Im}~{i\omega\over 8 m^2_p}\int dx^0 e^{-i\omega x^0}\left[
\omega^2 \epsilon^*_{ab}\epsilon_{cd} \langle T(Q^E_{ab}(0) Q^E_{cd}(x^0)) \rangle \right. \nonumber \\  & & + \left.
({\bf k}\times \epsilon^*)_{ab}  ({\bf k}\times \epsilon)_{cd} \langle T(Q^B_{ab}(0) Q^B_{cd}(x^0))\rangle\right],
\end{eqnarray}
where $({\bf k}\times\epsilon)_{ab}={{\bf \epsilon}_{acd}} {\bf
k}_c\epsilon_{db}$, and $\epsilon_{ab}$ the polarization tensor of the graviton. For non--rotating bodies the $SO(3)$ symmetry of the problem fixes the two--point function\footnote{Notice this is in the spirit of the ADS/CFT correspondence where isometries of the gravity side are mapped into global symmetries of the wordline theory \cite{nrgr2}}, 
\begin{equation}
\label{eq:g2pt}
\int dx^0 e^{-i\omega x^0} \langle 0| T Q^E_{ab}(0) Q^E_{cd}(x^0)|0\rangle = -{i\over 2}Q_{abcd}F(\omega),
\end{equation}
where the expectation value is taken in the vacuum state $|0\rangle$ of the internal theory, that is the vacuum state in the theory of the gapless modes of the BH responsible for absorption. We also introduce
\begin{equation}
Q_{abcd} = \left[\delta_{ac}\delta_{bd} +\delta_{ad}\delta_{bc} -{2\over 3}\delta_{ab}\delta_{cd}\right], 
\end{equation}
the projector onto symmetric and traceless two--index spatial tensors. 
The total cross section reads (after averaging over polarization)     
\begin{equation}
\sigma^{eft}_{abs}(\omega)={\omega^3\over 2 m^2_p}\mbox{Im}  F(\omega).
\end{equation}

The next step consists on matching this EFT result with the expression in the full theory for the absorption cross section of spinless BHs in the low frequency approximation. The result is \cite{starobinski,page},
\begin{equation}
\label{eq:gcross}
\sigma^{full}_{abs}(\omega)={1\over 45} 4\pi r^6_s\omega^4,
\end{equation}
with $r_s = 2G_Nm$, the Schwarzschild radius. We have then \cite{nrgr2}
\begin{equation}
\mbox{Im}  F(\omega) = 16 G^5_N m^6 |\omega|/45.  
\end{equation}

Here we used the equality of the magnetic and electric correlators, which follows from the duality invariance $E_{ab}\rightarrow -B_{ab}$, $B_{ab}\rightarrow E_{ab}$ of the linearized perturbation equations in a Schwarszchild background (see appendix).

\begin{figure}[t!]
    \centering
    \includegraphics[width=6cm]{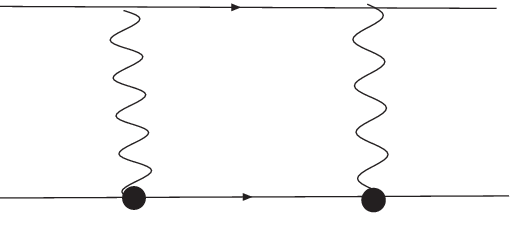}
\caption{Leading order contribution to the absorptive potential. The dots correspond to insertions of the leading worldline multipole operators.}
\label{figab} 
\end{figure}

Once the matching is obtained we can apply the formalism to a more complicated kinematical situation, such as a binary system composed of two BHs.  From (\ref{eq:g2pt})-(\ref{eq:gcross})  we see that $\langle Q^{E(B)}_{ab} Q^{E(B)}_{cd}\rangle\sim G^5_N m^6\omega^2$ and therefore, in the non--relativistic limit, with $\omega\sim v/r$,  $Q_{ab}^{E(B)}\sim L v^4/m_p$. Using the power counting techniques developed in \cite{nrgr,nrgr4} we will have for potential--gravitons \cite{nrgr2}
\begin{equation}
\int d\tau Q^E_{ab} E^{ab}[H]\sim v^{13/2},
\end{equation}
and similarly for the magnetic operator. By rotational invariance, and that fact that $Q^E_{ab}$ is traceless, we have $\langle Q^E_{ab}\rangle=0$ and the leading order absorptive contribution to $S_{eff}[x_a]$ is from the box diagram in Fig. \ref{figab},  with two insertions of the operators in (\ref{eq:BHgrav}) and two insertions of the leading order Newtonian interaction, $\int dx^0 \frac{m}{2m_p} H_{00}$ \cite{nrgr}, from the second particle. Since the magnetic operator does not couple to the latter, the leading contribution comes solely from the electric piece. 

The box diagram contributes to $\mbox{Im}S_{eff}$ a term that is of order $Lv^{13}$ for non--spinning BHs, $v^8$ order higher than the leading quadrupole radiation \cite{poissonab1}.  However, for spinning BHs this effect may be enhanced \cite{Tagoshi,poissonab2}, as we show later. 
Explicitly calculating the diagram of Fig. \ref{figab} we find 
\begin{eqnarray}
\nonumber
iS_{eff}[x_1,x_2] &=& {m^2_2 \over 8 m^4_p} \int {dx^0_1}  {d{\bar x}^0_1} {dx^0_2}  {d{\bar x}^0_2}   \langle T H_{00}(x^0_2) E_{ij}(x^0_1)\rangle  \langle T H_{00}({\bar x}^0_2) E_{rs}({\bar x}^0_1)\rangle\\
& &{} \times \langle T Q^E_{ij}(x^0_1) Q^E_{rs}({\bar x}^0_2)\rangle+(1\leftrightarrow 2) + \cdots.
\end{eqnarray}
We also have
\begin{equation} 
\langle T H_{00}(x^0,{\bf x}) E_{ij}(0) \rangle = -{i\over 16\pi} \delta(x^0) \partial_i \partial_j {1\over |{\bf x}|},
\end{equation}
and  defining $q^{a}_{ij}(t)=\partial^a_i\partial^a_j \frac{1}{r} = \frac{1}{r^3}(\delta_{ij}-3n_in_j)$ for $(a=1,2)$, we obtain \cite{nrgr2}
\begin{equation}
S_{eff}[x_1,x_2]= {1\over 2} G^2_N\left(\sum_{a\neq b} \int {d\omega\over 2\pi} F_a(\omega)m^2_b |q^{a}_{ij}(\omega)|^2 \right)+\ldots ,
\end{equation}

Recall that the imaginary part of the effective action measures the total number of gravitons emitted by a given configuration $\{x_\mu^a\}$ over an arbitrarily large time $T\rightarrow\infty$,
\begin{equation}
{1\over T} \,\mbox{Im} S_{eff}(x_a) =  {1\over 2} \int dE d\Omega
{d^2\Gamma\over dE d\Omega},
\end{equation}
where $d\Gamma$ is the differential rate for graviton emission
from the binary system from which the power spectrum is computed,
\begin{equation}
P= \int dE d\Omega E \frac{d^2 \Gamma}{d\Omega dE}\label{pws}.
\end{equation}

Therefore, using (\ref{pws}) the binding energy loss due to absorption is given by
\begin{equation}
P_{abs} = {16\over 45} G^7_N \left\langle\sum_{a\neq b} m^6_a m_b^2 {\dot q}^{a}_{ij} {\dot q}^{a}_{ij} \right\rangle.
\end{equation}

If we particularize it now for the case of circular orbit and $m_1 \ll m_2$ we obtain
\begin{equation} 
P_{abs}={32\over 5} G_N^7 m_2^6 m_1^2 \left\langle {{\bf v}^2\over r^8}\right\rangle,
\end{equation}
which agrees with the results in \cite{poissonab1,Tagoshi} for a test particle orbiting a Schwarzschild BH. Although we have considered the case of BHs, the methods generalize easily to NSs. In the latter the power spectrum for absorption of gravitational energy over an observation time $T$ is given by \cite{nrgr2}
\begin{equation}
\label{eq:general}
{dP_{abs}\over d\omega} = {1\over T} {G_N\over 32\pi^2}\left\langle \sum_{a\neq b} {\sigma^{a}_{abs}(\omega)\over\omega^2} m^2_b |q^{a}_{ij}(\omega)|^2 \right\rangle,
\end{equation}
where $\sigma^{a}_{abs}(\omega)$ is the graviton absorption cross section for each NS in the system. This is a very useful expression. Assuming a theoretical model of NSs is given, from which  $\sigma_{abs}(\omega)$ can be calculated, by measuring the absorption power spectrum we will gain knowledge about the internal structure of the NS and be able to rule out, or confirm, different models.   

\section{Absorption effects for spinning binary Black Holes}

Let us now add spin. Rotation introduces a few subtleties into the game, perhaps most remarkably the issue of superradiance. Among the first to propose such phenomena was Zel'dovich \cite{zeld}, who showed that amplification can occur for electromagnetic waves scattering off a rotating cylinder. In the case of a Kerr BH it was first noticed by Misner \cite{misner}. Also Bekenstein was able to show that the area law in BH dynamics implies that waves can be amplified by scattering off a rotating hole (see \cite{beke} and references therein). The mass of the BH decreases as the wave scatters off if the condition
\begin{equation}
{\tilde \omega} = \omega-m_l\Omega < 0\label{super}
\end{equation}
is satisfied. Here $\omega$ is the frequency of the incoming wave (hence $\tilde\omega$ in the ``co-rotating frame"), $m_l$ the azimuthal angular momentum number of the incoming wave with respect to the axis of rotation and $\Omega$ the rotational angular velocity. Therefore, the wave is amplified  at the cost of the energy of the BH. Heuristically, if the incoming wave is described in the $m_l$ mode by 
\begin{equation}
\Psi \sim e^{i(m_l\phi-\omega t)},
\end{equation} 
in the co-rotating frame ${\tilde \phi}=\phi-\Omega t$ and we have a phase shift, $m_l{\tilde\phi}-{\tilde\omega} t$. This means superradiance is somehow associated to ``negative" energies, as in the Penrose effect for the case of a Kerr BH\footnote{Notice that superradiance was first discovered in the realm of electromagnetism \cite{zeld}, and shows up in many other examples \cite{beke}, without any mention to ergospheres \cite{mtw} nor GR. Nevertheless, the equivalence principle tell us acceleration is tightly related to gravity. Indeed, the rotating frame was used by Einstein himself to argue about the geometry aspects of the gravitational field \cite{mtw}.} \cite{penrose}. From a more formal point of view it was shown for instance by Unruh that the solution to the wave equation in a Kerr background has ``incoming" modes with negative group velocity regardless of the value of $\omega$, whereas their phase velocity becomes positive in the superradiance regime \cite{unruh}. 

The superradiance effect bears in a delicate point once the optical theorem is applied, since the balance of energy for the BH renders
\begin{equation}
\frac{dE}{dt} = \frac{d}{dt} (E_{abs} - E_{amp}),
\end{equation}
with $E_{amp}$ the energy given away into the wave $(dm < 0)$ by superradiance, whereas $E_{abs}$ stands for the energy absorbed into the BH as in the non--rotational case.  If we define $\Delta E \equiv E_{abs}-E_{amp}$, it is clear that its sign is not predetermined since it is built up from different values of $m_l$, and some will obey (\ref{super}) and show superradiance. From Page's calculation \cite{page} we notice that is actually the case and the BH absorption ``probability" for gravitons is given in the low frequency approximation by
\begin{equation}
\Gamma_{s=2,\omega,l=s,m,p}=\frac{16}{225}\frac{A}{\pi} m^4[1+(m_l^2-1)a^2_{*}][1+(\frac{m_l^2}{4}-1)a^2_{*}]\omega^5(\omega-m_l\Omega)\label{astar}
\end{equation}
with $\Omega \sim {a_{*} \over 4 m}$, the rotational angular velocity, $a_{*} \equiv \frac{a}{M} = \frac{S}{m^2}$  and $A$ the area respectively in $G_N=c=1$ units, such that $S \sim m^3 \Omega$ as expected. In this expression $p$ represent the polarization of the incoming wave and we have kept only the dominate $l=s$ mode of the spheroidal harmonic for massless gravitons, e.g. $s=2$ \cite{page}. Notice that the superradiance term is manifest and implies a negative absorptive ``probability". The superradiance contribution averages out in the total cross section and we could therefore match for the latter without worrying about negative cross sections. However, the leading order absorption due to spin for a binary system is enhanced by three powers of the relative velocity precisely by this piece we need to handle \cite{Tagoshi,poissonab2}. The procedure will mimic the same steps as before. We generalize the two point function in (\ref{eq:g2pt}) to include the spinning part, and instead of matching for the total cross section we choose to scatter polarized spherical waves and obtain the spin component of the two point function by matching with the positive absorptive mode $m_l=-2$. Notice that matching is a theoretical procedure which does not rely necessarily on a priori ``observable" quantities. We could even match off--shell if necessary. The virtue of the method is that, once the matching is set, true predictive power is gained.

Let us proceed as we did before\footnote{Here we consider the $e^a_\mu$ local frame to be parallel transported along the worldline, as in the non--spinning case, and therefore the BH rotates with respect to this basis. We could have as well considered a co--rotating frame.}. Using the symmetries of the problem we get for the spinning part of the two point function,
\begin{equation}
\label{absp2} \int dx^0 e^{-i\omega x^0} \langle T Q^{E(B)}_{ij}(0)
Q^{E(B)}_{kl}(x^0)\rangle_{spin} = -{i\over
2} S_{ijkl} F_s(\omega)  
\end{equation}
where the frequency dependence in $S_{ijkl}$ enters in the spin and we have  
\begin{equation}
S_{ijkl} = \left[\delta_{ik}S_{jl} +\delta_{il}S_{jk}+\delta_{jl}S_{ik}+\delta_{jk}S_{il} \right] (1+\alpha{\bf S} ^2+\ldots ) + \ldots , 
\end{equation} 
with $\alpha$ a matching coefficient and the ellipses represent higher order corrections in spin. For instance we could add a term of the form ($S^{ij} = \epsilon^{ijk}S^k$)
\begin{equation}
\left[ \epsilon_{ikn}S_n S_jS_l + \epsilon_{jkn}S_n S_iS_l +
\epsilon_{iln}S_n S_jS_k + \epsilon_{jln}S_n S_kS_i
\right]F_{s^3}(\omega),
\end{equation}
which we can show that it corresponds to subleading effects in the small frequency approximation. 

There is yet another subtle point regarding the vacuum structure for the two point function in (\ref{absp2}). In the non--rotating case the leading order absorptive correction in the binary system comes from the box diagram since, by rotational invariance, the expectation value of the electric and magnetic quadrupole are set to zero. In the spinning case we know the Kerr solution is not rotationally invariant and presents a mass quadrupole \cite{thorne}, which is proportional to $S^2$. Therefore we would expect 
\begin{equation}
\langle Q^E_{ij}\rangle_{spin} \sim (S_{ik}S_{kj} - \frac{2}{3}\delta_{ij} {\bf S} ^2). 
\label{vev}
\end{equation}
For the case of a Kerr BH  we still have $\langle Q^B_{ij}\rangle =0$, and therefore the background breaks the `electric--megnatic' duality. Even though the term in (\ref{vev}) contributes to graviton scattering off the Kerr BH (the $E$ field have non--linear terms in the metric), it is easy to see it maps to subleading contributions. This follows from the fact that it is quadratic in spin and hence it matches onto terms which are suppressed in the small frequency expansion. Therefore, technically speaking we should treat the expectation value of the electric quadrupole as a `vev' (vacuum expectation value) and expand around it like we do for instance in the case of spontaneous symmetry breaking with a Higgs \cite{peskin}. All this means that the two point function in (\ref{absp2}) actually represents the time order product of the quadrupole fluctuations, $Q^E_{ij}-\langle Q^E_{ij}\rangle_{spin}$ for the electric piece, whose expectation value in the ``true" vacuum is zero \footnote{Notice that otherwise (\ref{absp2}) would signal the failure of the cluster decomposition for large time separation in the fields arguments, since we would expect the two point function to factorize, whereas it shows a linear behavior rather than a fourth power one.}.

Back to the absorption cross section.  The calculation follows in the same fashion as in the non--rotating case with the exception that, for the matching, we consider scattering of circularly polarized spherical waves so that we have two possible states, right $e_R = \frac{1}{\sqrt{2}}(e_+ + i e_{\times})$, and left $e_L = \frac{1}{\sqrt{2}}(e_+ - i e_{\times})$, movers \cite{mtw}. Here we deal with the $l=s=2,m_l=-2$ mode (no superradiance) given by $e_L$ spinning in the opposite direction of the rotation of the hole, located at the origin with spin aligned along the $z$-direction.  In the case of spinless BHs, matching is attained by comparison with the total cross section for plane waves. This follows from (\ref{astar}), summing over $m_l$ and keeping the leading order $l=2$ contribution, with the main difference being a flux normalization factor of $\pi \over \omega^2$ \cite{page}. However, for a rotating BH, in order to obtain the cross section we would need to consider the multipole expansion of the field in terms of spherical waves \cite{ryan} 
\begin{equation}
\psi^s_{lm_l} = Y^s_{lm_l}(\theta,\phi) \frac{1}{\sqrt{2\pi\omega}}\frac{e^{i(kr-\omega t)}}{r},
\end{equation}
with  $Y^s_{lm_l}(\theta,\phi)$ the spheroidal harmonics \cite{thorne}. The EFT formalism would stay the same, however, to make contact now with the more ``traditional" approach, we can decompose  a plane wave with helicity $p=-2$ moving in the $\hat {\bf z}$ direction, in terms spherical modes with coefficients $c^{s=2}_{l,m_l=-2} = i^l \sqrt{2l+1}$. If we keep only the dominant harmonic mode $l=2$, this effectively entails a factor of $\sqrt{5}$ for each external state in the optical theorem, and therefore an overall factor of $5$ multiplying (\ref{astar}). Notice this is also consistent with the result for non--rotating BHs. Henceforth, in the full theory side we get, again in the low frequency approximation,
\begin{equation}
\sigma^{full}_{abs}(\omega) = \frac{4\pi}{45}r_s^5\omega^3(1+3a^2_{*})+\ldots \label{astarcr},
\end{equation}
meanwhile from the EFT side we have
\begin{equation}
\sigma^{eft}_{abs}(\omega) =  G_N m^2 a_{*}(1+G_N^2m^4\alpha a_{*}^2){\omega^3\over
m^2_p} F_s(\omega)+ \ldots = \frac{\omega^3}{m_p^2} {\tilde F}_s(\omega) + \ldots ,
\label{sigeft}
\end{equation}
for the polarized absorption cross section, where we used $\bf {k} = \omega \hat {\bf z}$, ${\bf S} \cdot \hat {\bf z} = a_{*}G_Nm^2$, also
\begin{equation}
{\tilde F}_s(\omega) = F_s(\omega) (1+\alpha G_N^2m^4a_{*}^2) G_N m^2 a_{*}, 
\end{equation}
and the electric--magnetic duality as before. Even though the rotating background explicitly breaks the symmetry, it is still possible to show that the {\it linearized} equations for the perturbations are invariant under duality transformations also for the Kerr BH (see appendix).\\ 
 
Notice that $F_s$ is actually independent of $\omega$, although this is only true at leading order. However, keep in mind the spin tensor is a time dependent function which prevents the two point function from {\it shrinking} to an instantaneous interaction. There is moreover one last important point behind this expression, and that is that the imaginary component comes from the polarization contraction with the antisymmetric spin tensor rather than from $F_s$, which stays real\footnote{Notice also that matching with linearly polarized waves will not succeed since the spin tensor necessarily flips $\times$ to $+$ and viceversa.}.

All we are left over to do now is to match the EFT result of (\ref{sigeft}) with the full theory side in (\ref{astarcr}), from which we get $\alpha = \frac{3}{m^4G_N^2}$ and 
\begin{equation}
F_s(\omega) = \frac{1}{4} \frac{\mbox{Im}F(\omega)}{G_N^2m^3\omega} = \frac{4}{45}G_N^3 m^3. 
\end{equation}

Let us power count the spin dissipative contributions as we did previously. From the above expression we have
\begin{equation}
 \langle Q^{E(B)}_{ab} Q^{E(B)}_{cd}\rangle_{spin}\sim \omega G^3_N m^3 (G_N m^2 a_{*})(1+3a_{*}^2), 
\end{equation} 
and therefore in the NR limit
\begin{equation}
\left(Q_{ab}^{E(B)}\right)_{spin} \sim \sqrt{a_{*}(1+3a_{*}^2)} L v^{5/2}/m_p,
\end{equation}
which implies, after matching into NRGR for potential--gravitons, a spin dependent operator scaling as
\begin{equation}
\int d\tau \left(Q_{ab}^E\right)_{spin} E^{ab}[H] \sim \sqrt{a_{*}(1+3a_{*}^2)} v^5,
\end{equation}
and thus a $v^{10}$ contribution to the box diagram. This represents an enhancement of three powers of the relative velocity with respect to the spinless scenario \cite{Tagoshi,poissonab2}.     
Our final task to obtain the power loss is to compute the imaginary part of the effective action stemming from the spin dependent component of the absorptive piece in the box diagram. 
The first lines of the calculation are identical, and we also ignore the magnetic piece since it does not couple at leading order. We then have
\begin{equation}
S_{eff}[x_1,x_2]=  G^2_N\left ( \sum_{a\neq b} \int {d\omega\over 2\pi} {\tilde F}_s^b m^2_a q^{b}_{ij}(\omega) {q^b}^{*}_{il}(\omega)s^{b}_{jl}(\omega)\right)+\cdots ,
\end{equation}
with $s_{ij} = \epsilon_{ijk} s_k$, and $s_k=\frac{S_k}{|{\bf S} |}$, the unit vector in the direction of spin. 
Notice once again that the imaginary part of this expression comes from the spin contraction (remember it is an antisymmetric tensor) rather than $F_s$. To obtain the absorption power loss due to spin we use (\ref{pws}) one more time. Therefore we need: the imaginary part in frequency space, a factor of $2$ due to the optical theorem and the $E=\omega$ extra piece. Garnering all these together we end up with
\begin{eqnarray}
\nonumber
P^{spin}_{abs} &=& {8\over 45} G^6_N  \left\langle \sum_{A\neq B} m^5_A m^2_B\left({\dot q}^A_{ij} q^A_{il}s^A_{jl}\right)\left(a^A_{*}+3 (a^A_{*})^3\right) \right\rangle\\ &=& -{8\over 5}G_N^6 m_1^2m_2^2\left\langle {\frac{{\bf l}\cdot{\bf \xi}}{r^8} }\right\rangle \label{pspinabs}
\end{eqnarray}
where ${\bf l} = {\bf r}\times{\bf v}$, with ${\bf v} =\dot{\bf r}$, and ${\bf \xi} \equiv \sum_A m_A^3{\bf s} _A\left(a^A_{*}+3 (a^A_{*})^3\right)$. If we consider the regime of a rotating test particle ($m_2,a_\star$) moving in a circular orbit on the equatorial plane of a Schwarzschild BH $m_1$, such that $m_2 \ll m_1$, we obtain from Eq. (\ref{pspinabs})
\begin{equation}
P^{spin}_{abs} = -\epsilon \frac{8}{5} \frac{G_N^6m_2^5m_1^2 \omega_2}{r^6} (a_{*}+3 a_{*}^3)\label{psab}
\end{equation}
where $\omega_2$ is the orbital frequency of the test particle ($\omega_2=v_2/r$), and $\epsilon = \frac{\bf l}{|\bf l|}\cdot {\bf s} $. This is in agreement with the results in \cite{Tagoshi,poissonab2}.

We can extend the results now to the case of NS binary systems. This is however a bit more involved than before. Imagine for instance we have a theory which describes a rotating NS and we are able to calculate an expression similar to (\ref{astar}) for the graviton $(s=2)$ absorption probabilities for a given $(l,m_l)$ spheroidal mode. Let us denote this probability as $\Gamma^{NS}_{l,m_l}[\omega,\Omega]$. We assume the NSs show also a superradiant regime, at least for very compact ones, and therefore we write
\begin{equation}
\Gamma^{NS}_{l,m_l}[\omega,\Omega] = \tilde\Gamma^{NS}_{l,m_l}[\omega-m_l\Omega,\omega] (\omega-m_l\Omega),
\end{equation}
with $\tilde\Gamma^{NS}$ an even function of $\omega-m_l\Omega$. We are not aware of whether this is a sensible assumption, since it would depend on the equation of state. Let us however proceed as a working hypothesis. Working in the small frequency approximation, where the EFT is valid, we have $m\omega<1$ , which for rapidly rotating NSs implies $\omega < m_l \Omega$. Also, since $l \ge 2$,  let us proceed as before and consider the $l=2$ dominante mode. In other words, we assume corrections scale as $(m\omega)^l$, which are suppressed for $l>2$. Within this regime we have
\begin{equation}
\Gamma^{NS}_{l,m_l}[\omega,\Omega] \sim -\gamma^{NS}_{m_l,l=2}[\omega]m_l \Omega\label{gamns}, 
\end{equation}
with $\gamma^{NS}$ the leading order term in the low frequency approximation. This expression is equivalent to (\ref{astar}), with $\omega < m_l\Omega$. We can now generalize the power loss due to absorption for spinning NSs (we removed the $NS$ label for simplicity),
\begin{equation}
{dP^{spin}_{abs}\over d\omega} = {1\over T} {5G_N\over 32\pi} \left\langle \sum_{a\neq b} \frac{\gamma_{-2}^b[\omega]a_{*}}{\omega^4} m^2_a q^{b}_{ij}(\omega) {q^b}^{*}_{il}(\omega)s^{b}_{jl}(\omega)\right\rangle+\ldots .
\label{newabsNS}
\end{equation}

This expression generalizes the result in (\ref{eq:general}), and can be similarly used to discern between different models for rotating NSs.

\subsection{A test spinning black hole in a background spacetime}

Another simple application of the results in the previous section is to consider the motion of a small test Kerr BH in a  background space time whose curvature length scale $\cal R$ is much larger than $r_s$. We proceed as in \cite{nrgr2} and compute the effective action $S_{eff}[x]$ ignoring gravitational radiation effects such that we can treat the background spacetime described by $E^{ab},B^{ab},$ as external fields.  The leading term arises from two insertions of the operators $Q^{E,B}_{ab}$ and reads \cite{nrgr2}
\begin{equation}
2~\mbox{Im}S_{eff}[x]= \mbox{Im}~i\int d\tau d{\bar \tau}\langle T Q^E_{ab}(\tau) Q^E_{cd}({\bar \tau})\rangle\left[E^{ab}(\tau) E^{cd}({\bar\tau})+B^{ab}(\tau) B^{cd}({\bar\tau})\right]+\ldots
\end{equation}
whose imaginary part enables us to compute the power loss absorbed by the BH by following the same steps as before. Calculating in the local frame we find

\begin{equation}
P_{abs} = {8\over 45}a_*(1+3a_*^2) G^4_N m^5\left \langle \left({\dot E}_{ij}E_{il}+  {\dot B}_{ij} B_{jl}\right)s_{jl}\right \rangle ,
\label{newabs}
\end{equation}
with $s_{ij}$ the unit spin tensor we defined before and $a_{*}$ describing the rotation of the hole.
Since we have assumed that the three-frame $e^a_\mu$ is parallel transported along the worldline, where the spin tensor is defined, we have ${\dot E}^{ij} = e^i_\mu e^j_\nu (v\cdot D) E^{\mu\nu}$. 

\section{Conclusions}

In this paper we generalized the worldline approach of \cite{nrgr2} to include spin. We encountered a few subtleties along the way. Particularly the issue of superradiance, which turned out to contribute the leading order effect for the dissipative power loss of non--relativistic spinning binary BHs. Remarkably, an enhancement of three powers of the relative velocity was shown, in agreement with earlier calculations for test particles orbiting a Kerr BH  \cite{Tagoshi,poissonab2}.  Similarly to what was done for non--rotating objects \cite{nrgr2}, we generalized the result to other spinning bodies, such as NSs, once a model of the NS is given from which the absorption probabilities can be computed. We also obtained the power loss due to absorption for a test spinning black hole in a background spacetime. To my knowledge, the expressions in (\ref{newabsNS}) and (\ref{newabs}) are new results which have not appeared in the literature until now. 
In order to obtain the results we relied on the electric--magnetic duality for the correlation function in the wordline theory. As shown in the appendix this follows from the properties of the Teukolsky equations \cite{teuk}. 

To obtain higher order corrections to the absorption cross section is just a matter of computational work. Once these are included, the issue of divergences re--appears. The latter however are easily tamed by the methods of \cite{nrgr,nrgr3}. Dissipation due to finite size effects, such as the self--induced quadrupole due to spin \cite{nrgr3}, can also be studied, although the effects are subleading. These computations are currently undergoing.
 
\vskip 1cm
 
\centerline{\bf Acknowledgments}

\vskip 0.5cm

I would like to thank Ira Rothstein for insightful comments and discussions during which parts of this work were clarified, and to Steve Giddings and Walter Goldberger for comments. I would like to thank also Ted Newman, Eric Poisson and Manuel Tiglio for valuable help on the subject of the appendix. This work was supported in part by the Department of Energy under Grants DOE-ER-40682-143 and DEAC02-6CH03000, by grant RFPI-06-18 from the Foundational Questions Institute (fqxi.org) and funds from the University of California.

\vskip 1cm

\centerline{\bf Appendix: ``Electric--Magnetic" duality}

\vskip 0.5cm

Here we will show how does the electric--magnetic duality arises for linearized perturbations around Schwarzschild and Kerr BHs, which allowed us to simplify the calculations and equate the correlation two point functions for the electric and magnetic quadrupoles around the vacuum, even when the latter breaks the symmetry. This is therefore a non trivial result given the non--linear structure of  GR. The first hint that such symmetry may be present is given by the fact that  In Schwarzschild, and also Kerr BHs,  the perturbations associated with $E_{ab}$ are of even parity, and those associated with $B_{ab}$ are of odd parity, and the reflection and transmission coefficients turn out to be equal, even though the effective potentials are different\footnote{I would like to thank Eric Poisson for pointing this out to me.} \cite{Chandra}. In this appendix we show the aforementioned duality, which turns out follows from the Teukolsky equations \cite{ teuk} using the Newman--Penrose (NP) formalism \cite{np}.

As it was shown in \cite{tedref} Einstein equations can be written in terms of a complex field, $Z_{ab}$, defined as
\begin{equation}
Z_{ab} = E_{ab} + i B_{ab} ~ \label{eb},
\end{equation}
which transforms as $Z_{ab} \to i Z_{ab}$ under duality, and therefore the equations are invariant. A similar result was shown in \cite{newman} for the case of gravitational perturbations around a flat background, where (\ref{eb}) represents the field of the perturbation. Notice that in the case of a non trivial background, where the duality may be `spontaneously broken', the residual duality at the level of the perturbations around it, is not guarantee due to the non--linear structure of GR.
Here we will show however how the same object as in (\ref{eb}) appears also for the case of a Kerr background extending the duality to other than flat space.\\

In \cite{teuk} Teukolsky showed that the linearized equations for the gravitational perturbation around a rotating BH can be written in terms of the NP quantities, $\psi_{0(4)}$, defined as
\begin{eqnarray}
 \psi_0 &=& C_{\alpha\beta\gamma\delta} l^\alpha m^\beta l^\gamma m^\delta , \\
 \psi_4 &=& C_{\alpha\beta\gamma\delta} n^\alpha m^{*\beta} n^\gamma m^{*\delta}, 
\end{eqnarray}
where  $C_{\alpha\beta\gamma\delta}$ is the Weyl tensor and  $l^\alpha, m^\beta, n^\gamma$ the NP null vectors. If we now express the NP quantities in terms of $E_{ab}$ and $B_{ab}$; choosing the unit timelike vector $v^a =\frac{1}{\sqrt{2}} (l^a-n^a)$ to define the electric and magnetic components of the Weyl tensor, the result reads
\begin{equation}
\psi_{\pm} = Z_{ab} m^a_{\pm}m^b_{\pm}\label{np}
\end{equation}
where we introduced the notation $\psi_{-}=\psi_4,~\psi_+=\psi_0$ and $m_+ = m, ~ m_-=m^*$. From here it is straightforward to conclude that the duality thus holds in the case of a BH background.
Let us remark that the manipulations performed so far are valid for any spacetime background of the type Petrov D, which follows from Teukolsky analysis \cite{teuk}. 

There is yet another way to look at the duality. Suppose we define a new transformation $E_{ab} \leftrightarrow B_{ab}$. As we know from electromagnetism, this is also a symmetry of Maxwell equations provided $t \to -t$. In what follows we will show that similar considerations apply to the gravitational case.

The $E \leftrightarrow B$  transformation can be expressed as
\begin{equation}
\psi_{\pm} \to i \psi_{\mp}^* ~ . \label{dual}
\end{equation}

Our task now has transformed into exploring whether the Teukolsky equations are invariant under (\ref{dual}) plus $t \to -t$.
More precisely we are interested in showing that, for a given frequency $\omega$ and mode $m$, the perturbations associated with the magnetic and electric components of the Weyl tensor have the same asymptotic behavior.   
Using the separation of variables advocated by Teukolsky one can write the solution in the standard form in terms of spheroidal harmonics, $\Phi_{\pm}(r,t,\theta,\phi) = R^{l,m}_{\pm}(r) {\tilde Y}^{l,m}_{\pm}(t,\theta,\phi)$ with  $ {\tilde Y}^{l,m}_{\pm}(t,\theta,\phi) = S^{l,m}_{\pm}(\theta, a\omega) e^{i(\omega t + m \phi)}$  \cite{teuk}, where we included the time dependence in the spheroidal harmonic. As shown in \cite{teuk} the Teukolsky equations are formally invariant under the transformation $\Phi_+ = \psi_0 \to \Phi_- = \rho^4 \psi_4$, $s\to -s$, with $s$ the spin weight, and $\rho = r-ia cos\theta$. This symmetry can be recast as \cite{Chandra}
\begin{eqnarray}
S^{l,m}_-(\theta,a\omega) &=& S^{l,m}_+( \pi - \theta, a\omega),\\
R^{l,m}_- (r) &=& \Delta^2 \left(R^{l,m}_+\right)^* (r) \label{aha}, 
\end{eqnarray}
with $\Delta = r^2+a^2-2Mr $.  

Another important property of the solutions is that $S^{l,-m}_{\pm} (\pi-\theta,a\omega) = S^{l,m}_{\pm} (\theta,-a\omega)$ or equivalently $S^{l,m}_{\pm} (\theta,a\omega) = S^{l,-m}_{\mp} (\theta,-a\omega)$, from which we conclude $\left({\tilde Y}^{l,m}_{\pm}\right)^* = {\tilde Y}^{l,m}_{\mp}$ and therefore we get the symmetry transformation
\begin{equation}
\Phi_- \to  \Delta^2 \Phi_+^* \label{aha2}
\end{equation}

Notice that in the asymptotic region $r \gg M$ one has $\rho^4 \sim \Delta^2$, and also that this symmetry transformation takes $\omega \to -\omega$ or equivalently $t \to -t$.  Therefore we conclude that the transformation in (\ref{dual}), plus $t \to -t$, compose a symmetry of the Teukolsky equation in the regime where we are interested in, and in particular the absorption probabilities  of the Kerr  BH will manifestly obey the duality we used in the paper.

\end{document}